\documentclass[referee]{raa}
\usepackage{graphicx,times}
\usepackage{natbib}
\usepackage{amssymb,amsmath}
\bibpunct{(}{)}{;}{a}{}{,}
\usepackage[a4paper=true,dvipdfm=true,pagebackref=true]{hyperref}
\hypersetup{pdftitle = The title of my PDF, pdfauthor = My name, pdfsubject= The subject, pdfkeywords = keyword1 keyword2 keyword3}
\hypersetup{colorlinks = true, linkcolor = green, anchorcolor = red, citecolor = blue, filecolor = red, pagecolor = red, urlcolor = red}

\begin{document}

\title{Mode coupling in solar spicule oscillations}

\volnopage{ {\bf 2012} Vol.\ {\bf X} No. {\bf XX}, 000--000}
\setcounter{page}{1}

\author{Z.~Fazel\inst{}   }

\institute{ Astrophysics Department, Physics Faculty,
University of Tabriz, Tabriz, Iran; {\it z$_{-}$fazel@tabrizu.ac.ir}  }


\abstract{In a real medium which has oscillations, the perturbations can cause the energy transfer between different modes.
The perturbation interpreted as an interaction between the modes is inferred as mode coupling.
Mode coupling process in an inhomogeneous medium such as solar spicules may lead to the coupling of kink waves to local
Alfv\'{e}n waves. This coupling occurs practically in any conditions when there is smooth variation in density in the radial
direction. This process is seen as the decay of transverse kink waves in the medium.
To study the damping of kink waves due to mode coupling, a 2.5-dimensional numerical simulation of the initial wave is
considered in spicules. The initial perturbation is assumed to be in a plane perpendicular to the spicule axis.
The considered kink wave is a standing wave which shows an exponential damping in the inhomogeneous layer after occurrence
of the mode coupling.
}

   \authorrunning{Z. Fazel }            
   \titlerunning{Mode coupling }  
   \maketitle

%
\section{Introduction}           
\label{sect:intro}

Spicules are one of the important phenomena in the solar chromosphere which are observed at the
limb (Zaqarashvili \& Erd\'{e}lyi~\cite{Tem2009}). In order to study the oscillations in spicules,
many works have been done observationally and theoretically (De Pontieu et al.~\cite{De2004}, Kukhianidze et
al.~\cite{Kukh2006}, Kuridze et al.~\cite{Kur2013}, Verth et al.~\cite{Verth2011}). The generation and propagation
of magnetohydrodynamic (MHD) waves in the interval of the spicule lifetime (about $5-15$ min.) can be detected
by spicule observations.

Helioseismology can determine the properties of the solar phenomena from the observed oscillations which was
originally suggested by Zaqarashvili et al.~\cite{Tem2007} for the chromospheric spicules (Verth et al.~\cite{Verth2011}). By
estimating the period of oscillations, two types of MHD waves are observed in spicules: kink waves (Nikolski \& Platova
~\cite{Nikolski1971}, De Pontieu et al.~\cite{De2007}, He et al.~\cite{He2009a}, Ebadi et al.~\cite{Ebadi2012}) and Alfv\'{e}n waves
reported by Jess et al.~\cite{Jess2009}. Since spicules are denser than surrounding
coronal plasma (Beckers~\cite{bec68}), they can be modeled as cool magnetic tubes embedded in hot coronal plasma.

Damping of MHD waves (kink or Alfv\'{e}n waves) can be considered as mechanisms of the solar coronal heating.
When MHD waves interact with plasma inhomogeneities, a number of physical phenomena is generated such as: resonant
absorption, mode coupling, phase mixing, and wave dispersion.

The conversion of energy from the incompressible kink mode to an Alfv\'{e}n wave describe mode coupling process
in an inhomogeneous medium. This process can take place in each oscillating phenomena. For example in sunspots,
it takes place between the p-mode seismic wave field of the solar interior and the oscillations in the overlying
atmosphere (conversion of fast to slow modes) (Cally et al.~\cite{Cally1994}). Pascoe et al.~\cite{Pascoe2010} have studied the
damping of kink waves due to mode coupling in solar coronal loops. They have demonstrated that the observed loop
displacements are the coupled kink-Alfv\'{e}n waves, i.e. transverse footpoint motions travel along the loop and
through the inhomogeneity at the loop boundary couple to Alfv\'{e}n waves. At the presence of an inhomogeneous
layer, the Alfv\'{e}n speed varies continuously and resonant absorption occurs where the phase speed of the kink
wave matches the local Alfv\'{e}n wave speed ($C_{k}= V_{A}(r)$) (Allan \& Wright~\cite{Allan2000}, Hood et
al.~\cite{Hood2013a}). Terradas et al.~\cite{Terradas2010} demonstrated that the damping of the transverse motions
through mode coupling is frequency dependent. Using the CoMP data (Tomczyk \& McIntosh~\cite{Tomczyk2009}),
Verth et al.~\cite{Verth2010} found evidence for this frequency
that strengthen the interpretation of the observed propagating Doppler shift oscillations as the coupled kink-
Alfv\'{e}n waves.

Here, we study the damping of the observed transverse oscillations of the solar spicule axis. This study is
referring to the results obtained by analyzing of the time series of Ca II H-line obtained from Hinode/SOT on
the solar limb Tsuneta et al.~\citep{Tsuneta2008}. These observed transverse oscillations were interpreted as standing kink
waves (Ebadi et al.~\cite{Ebadi2012}). This paper is organized as follows. The basic equations and theoretical model are
presented in section $2$. In section $3$, numerical results are presented and discussed, and section $4$
contains the conclusion.

\section{Basic equations of model}

The damping of standing kink waves is studied in a spicule environment through the mode coupling mechanism.
We perform a 2.5-dimensional simulation of MHD equations in a stratified medium. Non-ideal MHD equations in
the plasma dynamics are as follows:
\begin{equation}
\label{eq:mass} \frac{\partial \mathbf{\rho}}{\partial t} + \nabla
\cdot (\rho \mathbf{V}) = 0,
\end{equation}
\begin{equation}
\label{eq:momentum} \rho\frac{\partial \mathbf{V}}{\partial t}+\rho(\mathbf{V} \cdot \nabla)\mathbf{V} =
-\nabla p + \rho\mathbf{g}+ \frac{1}{\mu_{0}}(\nabla \times \mathbf{B})\times\mathbf{B}+ \rho\nu\nabla^2\mathbf{V},
\end{equation}
\begin{equation}
\label{eq:induction}
\frac{\partial \mathbf{B}}{\partial t} = \nabla\times(\mathbf{V} \times \mathbf{B}), ~~~\nabla \cdot \mathbf{B} = 0,
\end{equation}
\begin{equation}
\label{eq:adia}
\frac{\partial p}{\partial t} +\nabla\cdot (p \mathbf{V})= (1-\gamma)p\nabla\cdot \mathbf{V}.
\end{equation}
where $p=\rho RT/\mu$ is the pressure for the perfect gas; $\rho$ is the plasma density, $\mu$
is the mean molecular weight, $\mu_{0}$ is the vacuum permeability, $\nu$ is constant viscosity coefficient,
$\gamma =5/3$ is the adiabatic index, $g$ is the gravitational acceleration. And we apply
$\rho\nu = 2.2\times 10^{-17}$$T^{5/2}$ kg m$^{-1}$ s$^{-1}$ for a fully ionized $H$ plasma (Priest~\cite{Prie1982}).

\subsection{The equilibrium state}
\label{sect:Obs}

Vectors $\mathbf{V}$ and $\mathbf{B}$ in Eqs.~\ref{eq:mass}-~\ref{eq:adia} are the velocity and magnetic field
which are defined as follows:
\begin{eqnarray}
\label{eq:perv}
  \textbf{V} &=& \textbf{v}_{0} + \textbf{v}, \nonumber\\
  \textbf{B} &=& \textbf{B}_{0} + \textbf{b}.
\end{eqnarray}
where $\textbf{v}_{0}=v_{0}\hat{k}$ and $\textbf{B}_{0}$ are the equilibrium velocity and magnetic field. In order
to have the kink waves coupled with Alfv\'{e}n waves, the equilibrium magnetic field is considered as
$\textbf{B}_{0}=B_{0}(\alpha\hat{j}+\hat{k})$ where $\alpha$ is a constant value ($\alpha <1$). In this case, the
y-component of the magnetic field is smaller than its z-component by the factor $\alpha$. Since the equilibrium
magnetic field is force-free, the pressure gradient is balanced by the gravity force via this equation
$-\nabla p_{0}(x,z) + \rho_{0}(x,z) \textbf{g}=0$, where $\textbf{g}$=$(0, 0, -g)$, and the pressure in an equilibrium state is:
\begin{equation}
\label{eq:presse}
 p_{0}(x,z)= p_{0}(x)~\exp(-z/H).
\end{equation}
Since coupling between kink and Alfv\'{e}n waves occurs practically in any conditions when there is smooth density variations in
the radial direction (here, x-direction) (Ruderman \& Roberts~\cite{Ruderman2002}, Soler et al.~\cite{Soler2011}), the density
profile is written in the following form:
\begin{equation}
\label{eq:density}
\rho_{0}(x,z)= \rho_{0}(x)~\exp(-z/H),
\end{equation}
where $H=\frac{RT}{\mu g}$ is the pressure scale height and the atmosphere is considered isothermal. And $\rho_{0}(x)$ is considered as
(De Moortel et al.~\cite{De1999}, Ruderman \& Roberts~\cite{Ruderman2002}, Karami \& Ebrahimi~\cite{Karami2009}):
\begin{equation}
\label{eq:densityx}
\rho_{0}(x)= \frac{1}{2} \rho_{0} [1+d_{\rho}-(1-d_{\rho})\tanh((x-1)/d)].
\end{equation}
here, $d_{\rho}= \frac{\rho_{e}}{\rho_{0}}= 0.01$ where $\rho_{0}$ is the plasma density in the spicule and $\rho_{e}$ is the
external density, and $d$ is the width of inhomogeneous layer.

\subsection{Perturbation Equations}

In order to make the continuous displacements of the tube axis of a spicule, the perturbation in the velocity and magnetic field
is considered at the lower boundary of the tube. Vectors $\mathbf{v}$ and $\mathbf{b}$ in Eq.~\ref{eq:perv} are the perturbed velocity
and magnetic field which are defined as $\textbf{v}=(v_{x}, v_{y}, v_{z})$ and $\textbf{b}=(b_{x}, b_{y}, b_{z})$, respectively.
The initial perturbed velocity is considered as two dimensional dipole form as introduced by Pascoe et al.~\cite{Pascoe2010}:
\begin{eqnarray}
\label{eq:shear field}
 v_{x}(x,z,t=0) &=& \frac{x^2-z^2}{(x^2+z^2)^2}, \nonumber\\
 v_{y}(x,z,t=0) &=& \frac{2xz}{(x^2+z^2)^2}, \nonumber\\
 v_{z}(x,z,t=0) &=&  A_{v}\sin(\pi x)\sin(\pi z).
\end{eqnarray}
and
\begin{eqnarray}
\label{eq:icv}
(b_{x}, b_{y}, b_{z})(x,z,t=0) &=& A_{b}\sin(\pi x)\sin(\pi z),\nonumber\\
p(x,z,t=0) &=& A_{p}\sin(\pi x)\sin(\pi z).
\end{eqnarray}
where $A_{v}$, $A_{b}$ and $A_{p}$ are the small amplitudes of the perturbed velocity, magnetic field and pressure (by choosing these
small amplitudes, the components tend almost to zero. These choices are made to avoid of some unwanted effects in our simulation code).
\\In order to see the variations of the perturbed velocity and magnetic field, linearized dimensionless
MHD equations with considered assumptions are as follows:
\begin{eqnarray}
\label{eq:mom}
\frac{\partial v_{x}}{\partial t}+ v_{0}\frac{\partial v_{x}}{\partial z} = \frac{1}{\rho_{0}(x,z)}(\frac{\partial b_{x}}{\partial z}-\alpha\frac{\partial b_{y}}{\partial x}-\frac{\partial b_{z}}{\partial z})+ \nu\nabla^2 v_{x}, \nonumber\\
\frac{\partial v_{y}}{\partial t}+ v_{0}\frac{\partial v_{y}}{\partial z} = \frac{1}{\rho_{0}(x,z)}\frac{\partial b_{y}}{\partial z}+
\nu\nabla^2 v_{y} \nonumber\\
\frac{\partial v_{z}}{\partial t}+ v_{0}\frac{\partial v_{z}}{\partial z} =-\beta\frac{\partial P}{\partial z}-g +\frac{\alpha}{\rho_{0}(x,z)}\frac{\partial b_{y}}{\partial z}+\nu\nabla^2 v_{z},
\end{eqnarray}
\begin{eqnarray}
\label{eq:indu}
\frac{\partial b_{x}}{\partial t} = \frac{\partial v_{x}}{\partial z}- v_{0}\frac{\partial b_{x}}{\partial z}, \nonumber\\
\frac{\partial b_{y}}{\partial t} = \frac{\partial v_{y}}{\partial z}- v_{0}\frac{\partial b_{y}}{\partial z}, \nonumber\\
\frac{\partial b_{z}}{\partial t} = \frac{\partial v_{z}}{\partial z}- v_{0}\frac{\partial b_{z}}{\partial z}.
\end{eqnarray}
where $\beta(z)= \frac{p_{0}(z)}{B_{0}^2/2\mu}$ is the ratio of gas pressure to magnetic pressure.
The boundary conditions for a standing formalism are considered $\mathbf{v}=0$ and $\textbf{b}=0$ at $z= 0, L$.
Hence by considering these conditions in Eq.~\ref{eq:adia}, we have $p=0$ at $z= 0, L$ (Ruderman \& Roberts~\cite{Ruderman2002}).

\section{Discussion}

In order to solve the coupled Eqs.~\ref{eq:mom}, and~\ref{eq:indu} numerically, the finite difference and the Fourth-Order
Runge-Kutta methods are used to take the space and time derivatives, respectively.
The implemented numerical scheme is using by the forward finite difference method to take the first spatial derivatives with
the truncation error of ($\Delta x$), which is the spatial resolution in the $x$ direction. The order of approximation for
the second spatial derivative in the finite difference method is $O((\Delta x)^2)$. On the other hand, the Fourth-order
Runge-Kutta method takes the time derivatives in the questions. The computational output data are given in $17$ decimal
digits of accuracy (Fazel \& Ebadi~\cite{Fazel2014}). We set the number of mesh-grid points as~$256\times256$. The time
step is chosen as $0.0001$, and the system length in the $x$ and $z$ dimensions (simulation box sizes) are set to be
($0$,$4$) and ($0$,$20$).

In considered spicule condition, the value of the all presented parameters are (Murawski \& Zaqarashvili~\cite{Murawski2010},
Ebadi et al.~\cite{Ebadi2012}, Fazel \& Ebadi~\cite{Fazel2013}):
$L=6000$~km (spicule length), $a=250$~km (spicule radius), $d= 0.2 a= 50$~km (inhomogeneous layer width),
$n_{e}=11.5\times10^{16}$~m$^{-3}$, $V_{A0}=50$~km/s, $v_{0}=25$~km/s, $B_{0}=1.2\times10^{-3}$~Tesla, $T_{0}=14~000$~K,
$p_{0}=3.7\times10^{-2}$~N m$^{-2}$, $\rho_{0}=1.9\times10^{-10}$~kg m$^{-3}$, $g=272$~m s$^{-2}$, $R=8300$~m$^{2}$s$^{-1}$k$^{-1}$
(universal gas constant), $H= 750$~km, $\mu=0.6$, $\mu_{0}=4\pi \times10^{-7}$~Tesla m A$^{-1}$, $\tau= a/C_{k}= 5$~s
(the period of oscillations), $A_{v}= A_{b}= A_{p} = 10^{-8}$ (dimensionless amplitudes of perturbed velocity, magnetic field and pressure).

Considering Eqs.~\ref{eq:mom}, and~\ref{eq:indu}, the y-component of the velocity and the magnetic field define the y-independent
Alfv\'{e}n waves. By rewriting the y-component of these equations and by combining them, we can obtain
\begin{eqnarray}
\label{eq:ind}
\rho_{0}[\frac{\partial^2 v_{y}}{\partial t^2}+ (\textbf{v}_{0}\cdot \nabla)\frac{\partial v_{y}}{\partial t}] =
\frac{1}{\mu{0}}[(\textbf{B}_{0}\cdot\nabla)^2 v_{y} - (\textbf{v}_{0}\cdot \nabla)b_{y}]+
\rho_{0}\nu\nabla^2\frac{\partial v_{y}}{\partial t}
\end{eqnarray}
This equation describes the damped Alfv\'{e}n waves that have velocity and magnetic field perturbations in the y-direction
and propagate along the equilibrium magnetic field. If for a moment we forget about $\nu$ and adopt a local analysis, the
dispersion relation is obtained as $\sigma^2 = \frac{k_{\|}^2 B_{0}^2}{\rho_{0}} = k_{\|}^2 V_{A}^2$,
where $k_{\|}= \alpha k_{y} + k_{z}$ is the parallel wave number and $V_{A}$ is the local Alfv\'{e}n velocity.
The second row of Eqs.~\ref{eq:mom}, and~\ref{eq:indu} describes the damped Alfv\'{e}n waves which are coupled with
the first and third rows of Eqs.~\ref{eq:mom}, and~\ref{eq:indu}. The equations tell us that a transfer of energy from
motions described by the variables ($v_{x},v_{z},b_{x},b_{z}$) to motions described by the variables ($v_{y},b_{y}$) can
occur in the considered MHD waves. The variables ($v_{x},v_{z},b_{x},b_{z}$) are damped due to the coupling
to Alfv\'{e}n waves. These results are demonstrated in the following figures for the perturbed velocity and magnetic field variations.

\begin{figure}
\centering
\includegraphics[width=8cm]{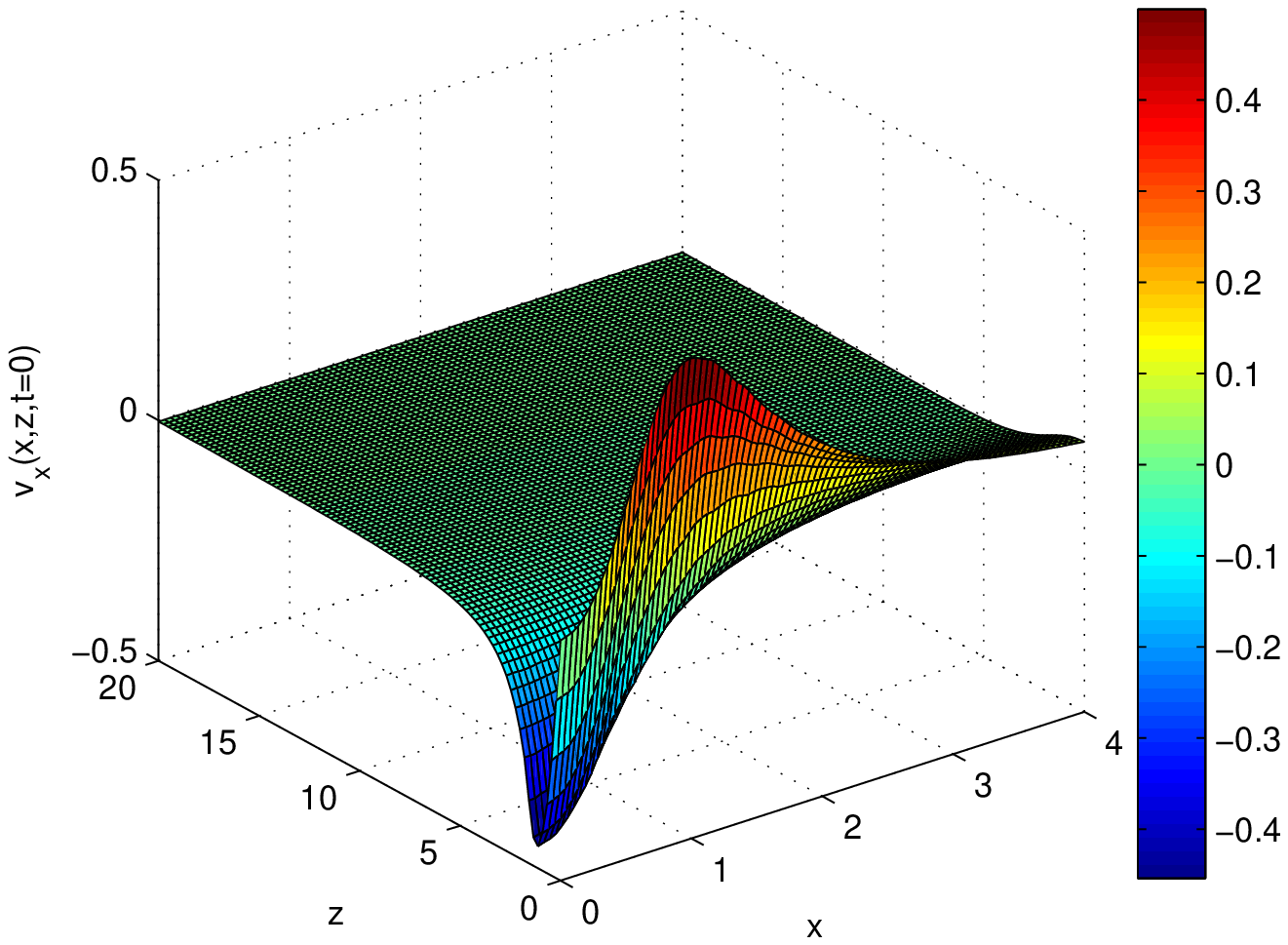}
\includegraphics[width=8cm]{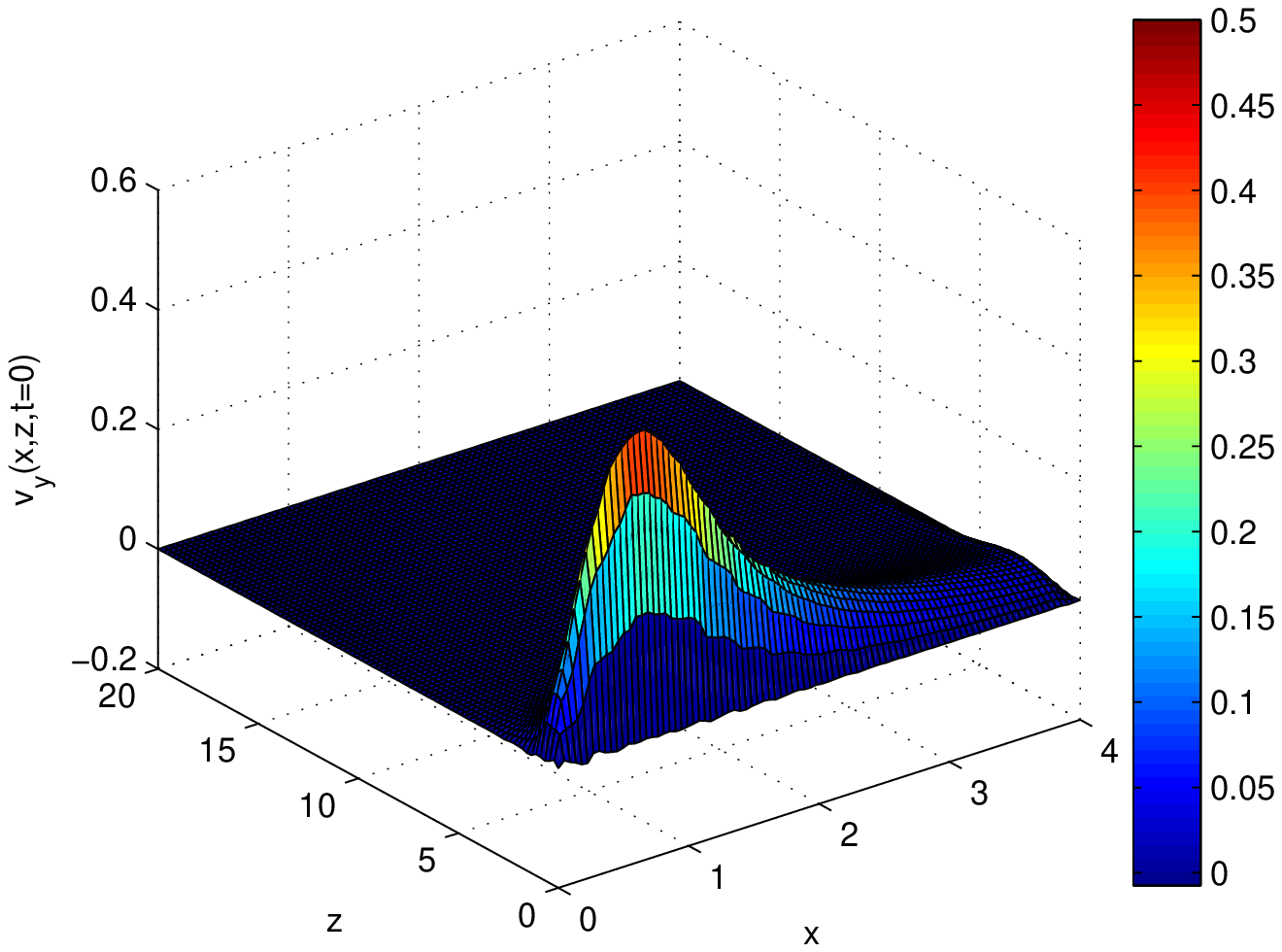}
\caption{The initial wave packet ($v_{x}$ and $v_{y}$) is showed at $t= 0$~s. \label{fig1}}
\end{figure}

At $t=0$, the initial perturbation is applied at lower z boundary. This perturbation propagates upward uniformly
and then it stops at upper z boundary (the standing case). Figure~\ref{fig1} is the plot the initial wave packet given by Eq.~\ref{eq:shear field}.
\\In the case of an inhomogeneous layer ($\rho_{0}/\rho_{e} > 1$), since the initial perturbation acts over all density regions (inside
the spicule, in the inhomogeneous layer and outside of spicule), the Alfv\'{e}n speed varies in these regions ($V_{A}(x,z)$). The
Alfv\'{e}n speed varies continuously in the inhomogeneous layer and resonant absorption occurs where the condition $v_{A} = v_{phase}$
is satisfied. Here, $v_{phase}= \omega/ k_{z}$, where $\omega$ is the angular frequency of kink oscillation and $k_{z}$ is the local
longitudinal wave number. In our simulation, the spicule is considered as a thin flux tube ($a= 250$~km radius) with a thin
inhomogeneous layer ($d= 0.2 a= 50$~km width). In this approach, the above condition is satisfied in the considered
inhomogeneous layer, so this is the subject of mode coupling to occur in the modeled spicule.

\begin{figure}
\centering
\includegraphics[width=8cm]{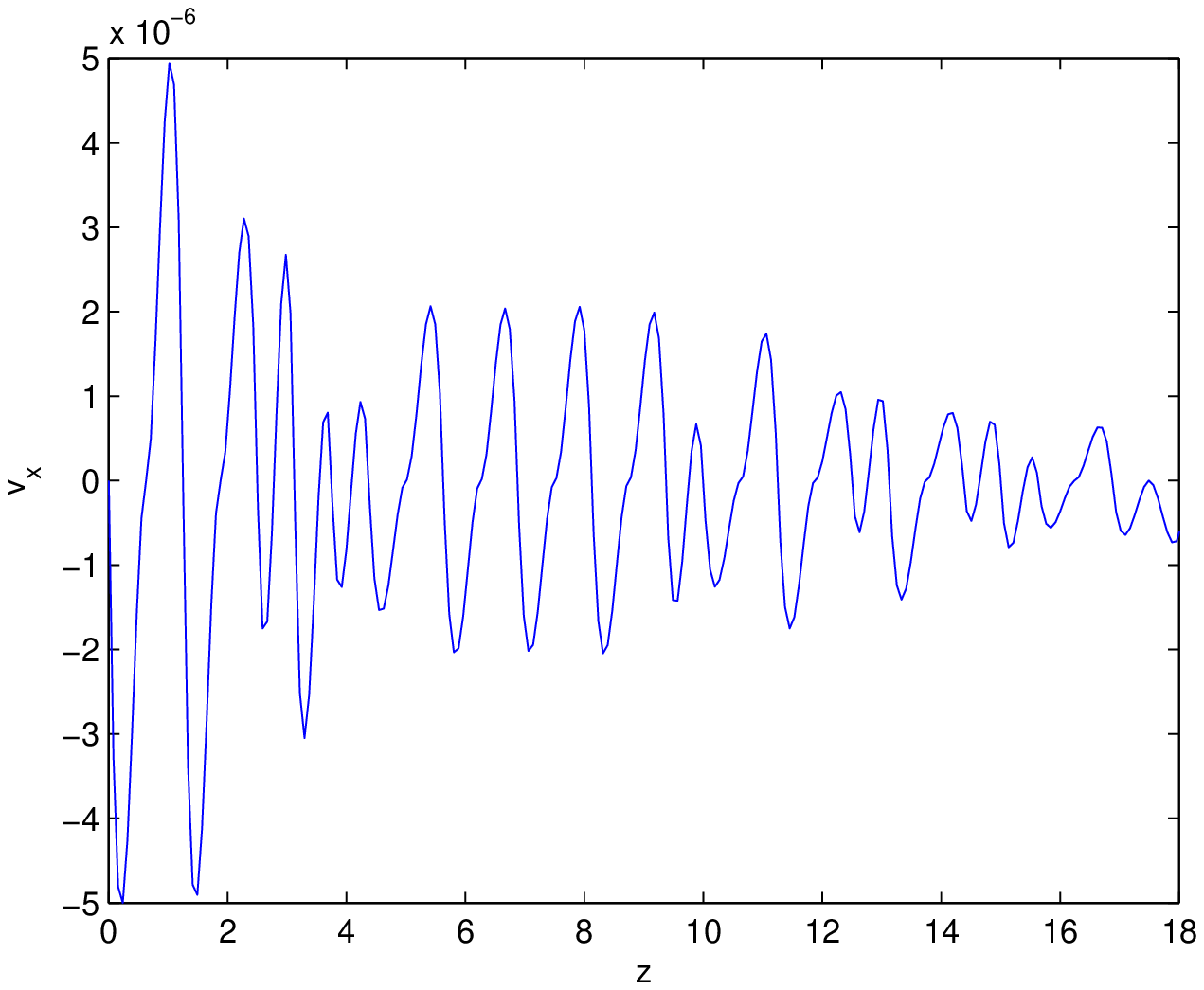}
\includegraphics[width=8cm]{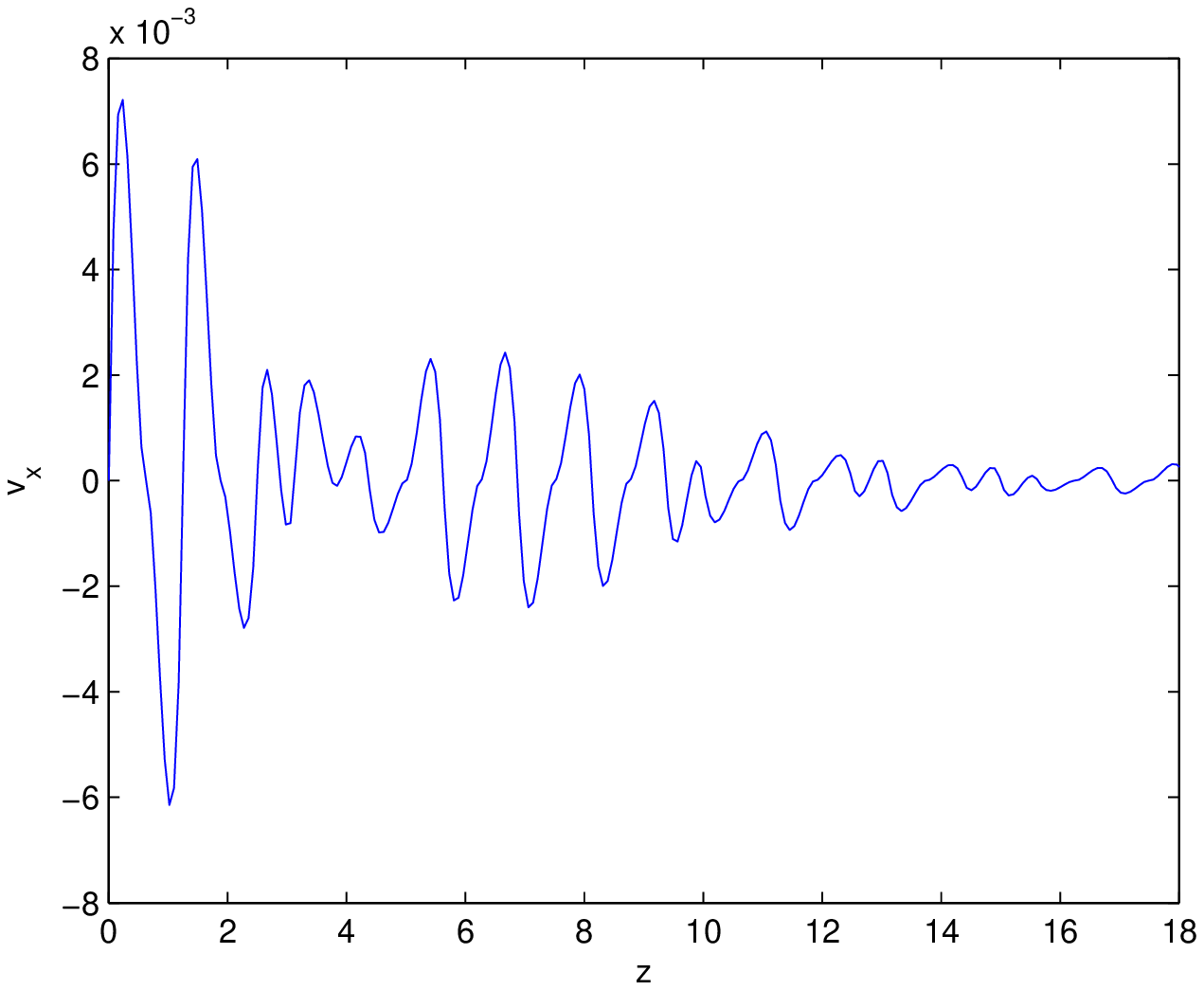}
\caption{The perturbed velocity variations are showed with respect to $z$ at two time steps: $t= 10 \tau$~s (top panel)
and $t=40\tau$~s (bottom panel). \label{fig2}}
\end{figure}

\begin{figure}
\centering
\includegraphics[width=8cm]{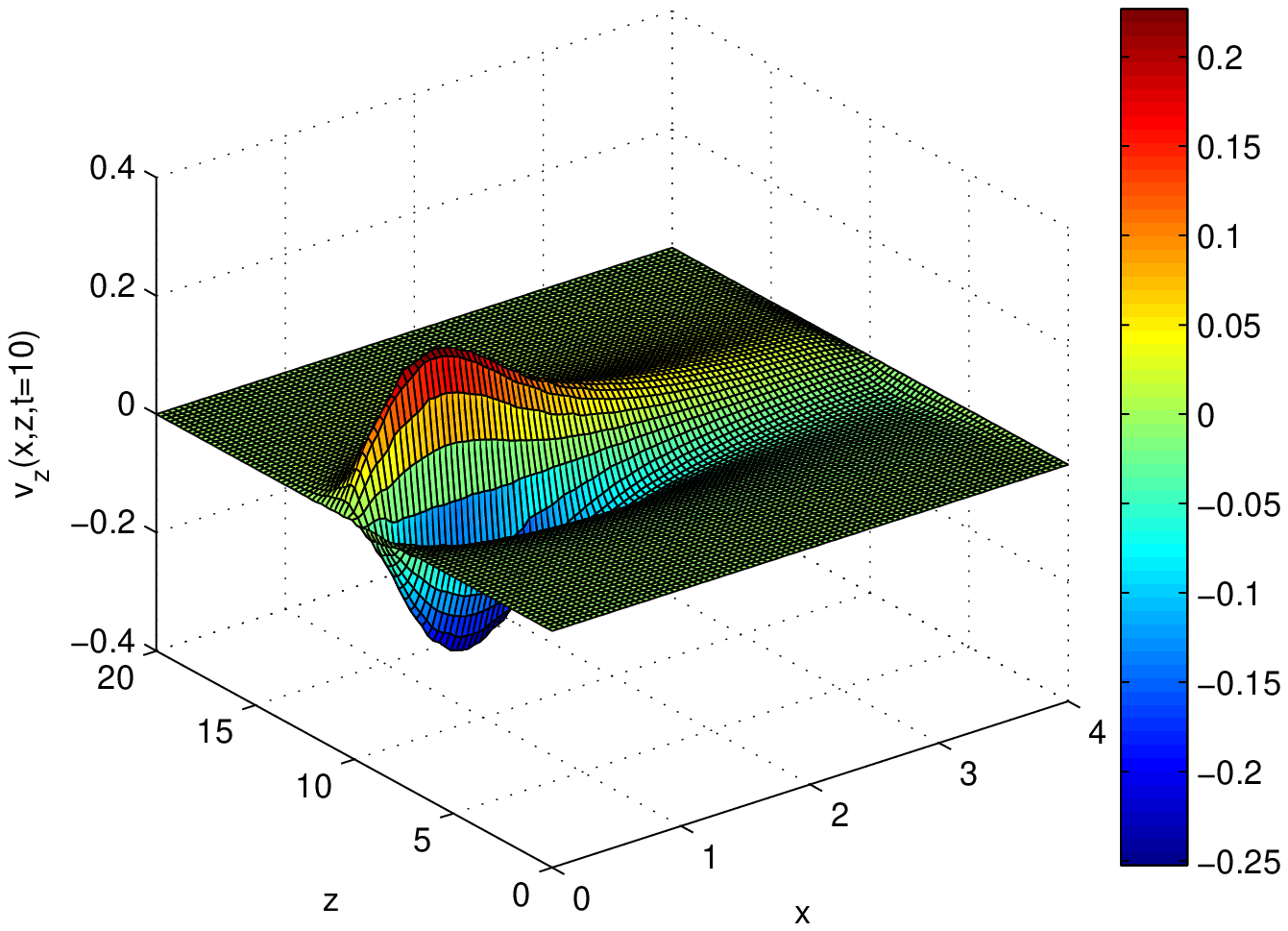}
\includegraphics[width=8cm]{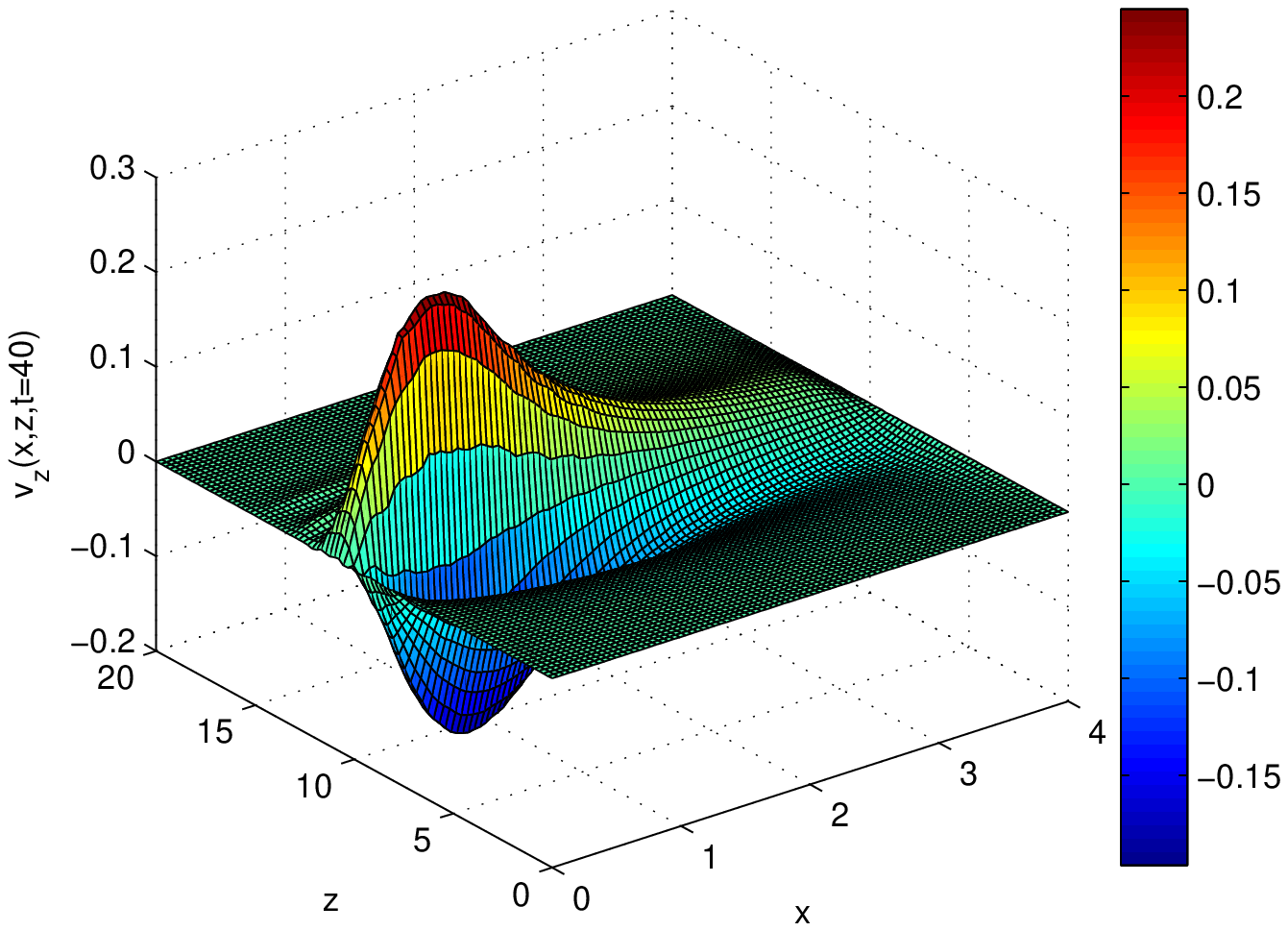}
\caption{The perturbed velocity in $x-z$ space is showed. Top panel corresponds to $t= 10 \tau$~s and the bottom to $t=40\tau$~s. \label{fig3}}
\end{figure}

Figure~\ref{fig2} shows the perturbed velocity variations, $v_{x}$ with respect to $z$ (height or propagation direction) at $t= 10 \tau$~s
(upper panel) and $t= 40 \tau$~s (lower panel), respectively. The perturbed velocity is normalized to $V_{A0}$.
At $t= 10 \tau$~s, the perturbed $v_{x}$ already shows the damping due to phase mixing occurring in the inhomogeneous layer.
By the later stage at $t= 40 \tau$~s, we see that the initial perturbation (the kink wave) has undergone the complete attenuation and
only the phase-mixed Alfv\'{e}n wave with bigger amplitude remains in the layer. The lower panel shows an exponential damping of $v_{x}$
with $z$. This is in good agreement with results obtained by Pascoe et al.~\cite{Pascoe2012}. They have demonstrated that for
standing kink modes with thin flux tube and thin boundary layer approximation, an exponential damping envelope is obtained.
In both panels at the first heights, total amplitude of velocity oscillations has values near to the initial ones.
As height increases, the perturbed velocity amplitude decreases to smaller values which demonstrates damping process due to mode coupling.
The coupling of the kink wave to a local Alfv\'{e}n mode causes a decrease in wave energy in the spicule, and so appears to damp the spicule
oscillation.
Figure~\ref{fig3} is the $3D$ plots of the perturbed velocity in $x-z$ space at two time steps (as mentioned in Figure~\ref{fig2}).

\begin{figure}
\centering
\includegraphics[width=8cm]{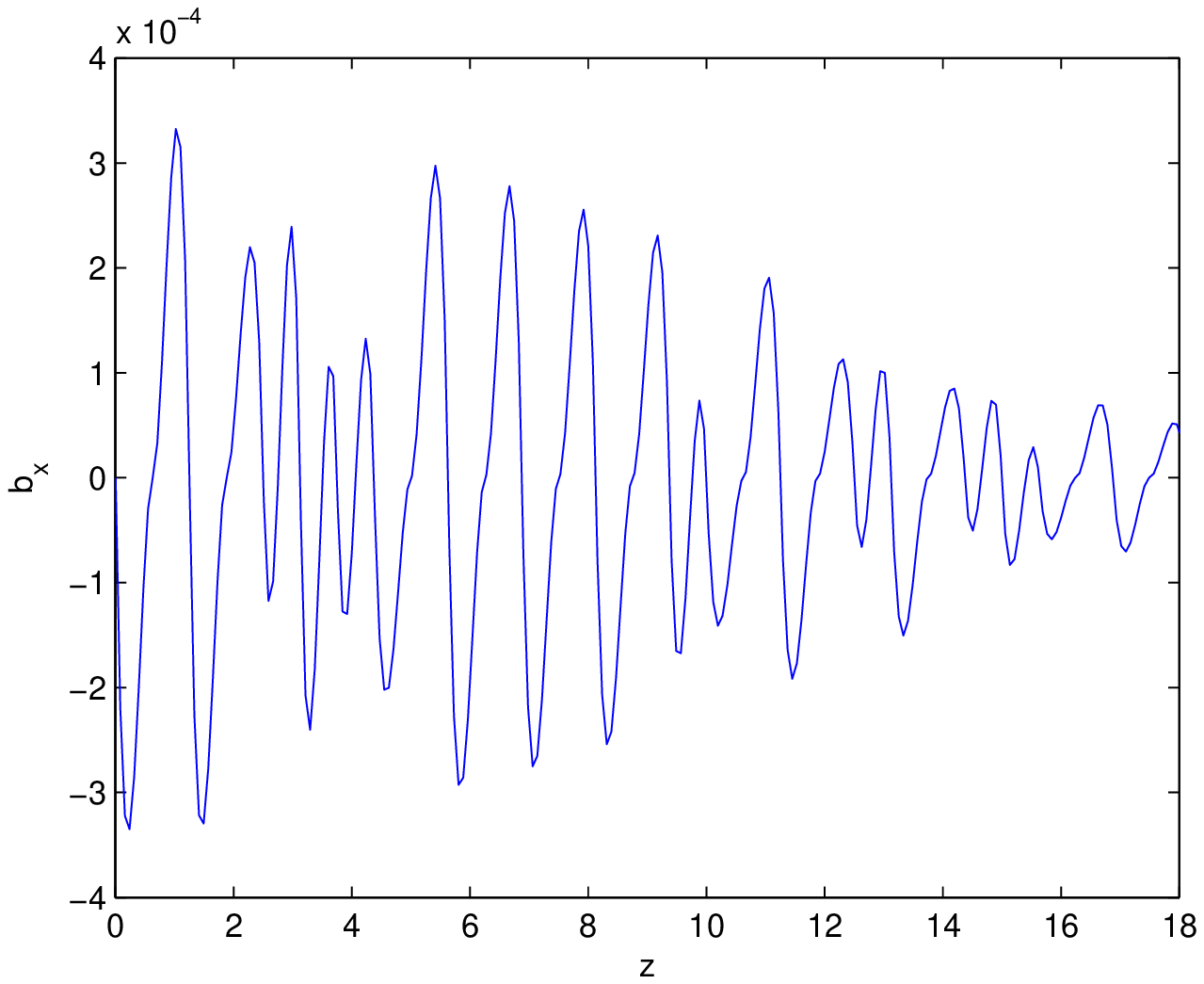}
\includegraphics[width=8cm]{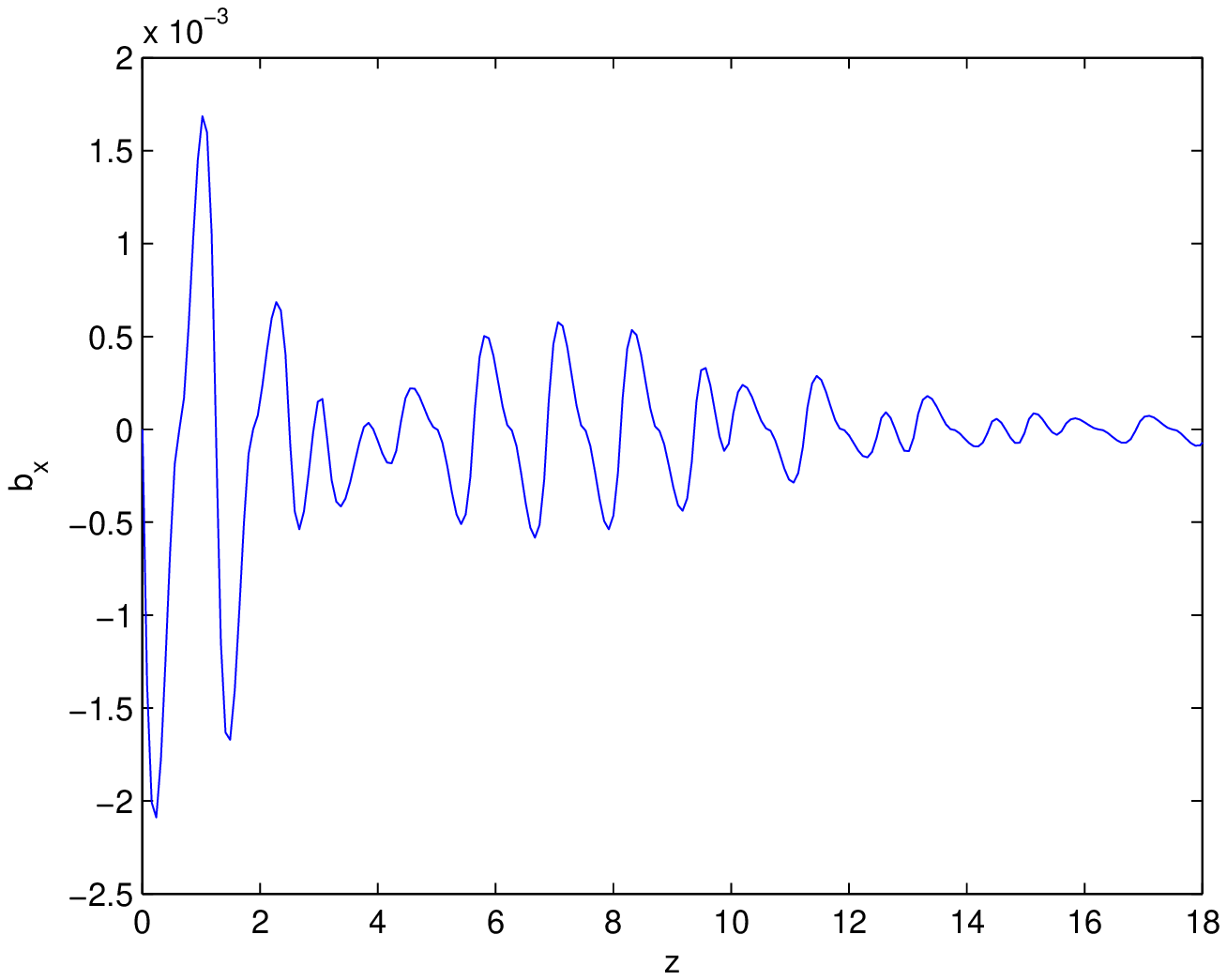}
\caption{The perturbed magnetic field ($b_{x}$) variations are showed with respect to $z$. Upper panel shows the kink mode along the
spicule axis at $x=250$~km. Lower panel shows the Alfv\'{e}n mode in the inhomogeneous layer at $x=475$~km. The perturbed magnetic
fields are normalized to $B_{0}$. \label{fig4}}
\end{figure}

Figure~\ref{fig4} demonstrates variations of the perturbed magnetic field component, $b_{x}$, with respect to $z$ at two locations:
spicule axis (according to our simulation box, the spicule axis is placed at dimensionless $x=2$ corresponding to $x=250$~km)
(upper panel), and inhomogeneous layer ($x=3.8$ corresponding to $x=475$~km) (lower panel).
The upper panel shows the upwards propagating kink oscillations (with smaller amplitude) along the spicule axis which undergoes
damping due to mode coupling as height increases. The lower panel shows the Alfv\'{e}n mode (with bigger amplitude) in the inhomogeneous
layer. The component $b_{x}$ grows in $z$ as energy is transferred to the Alfv\'{e}n mode from the kink mode.

\begin{figure}
\centering
\includegraphics[width=8cm]{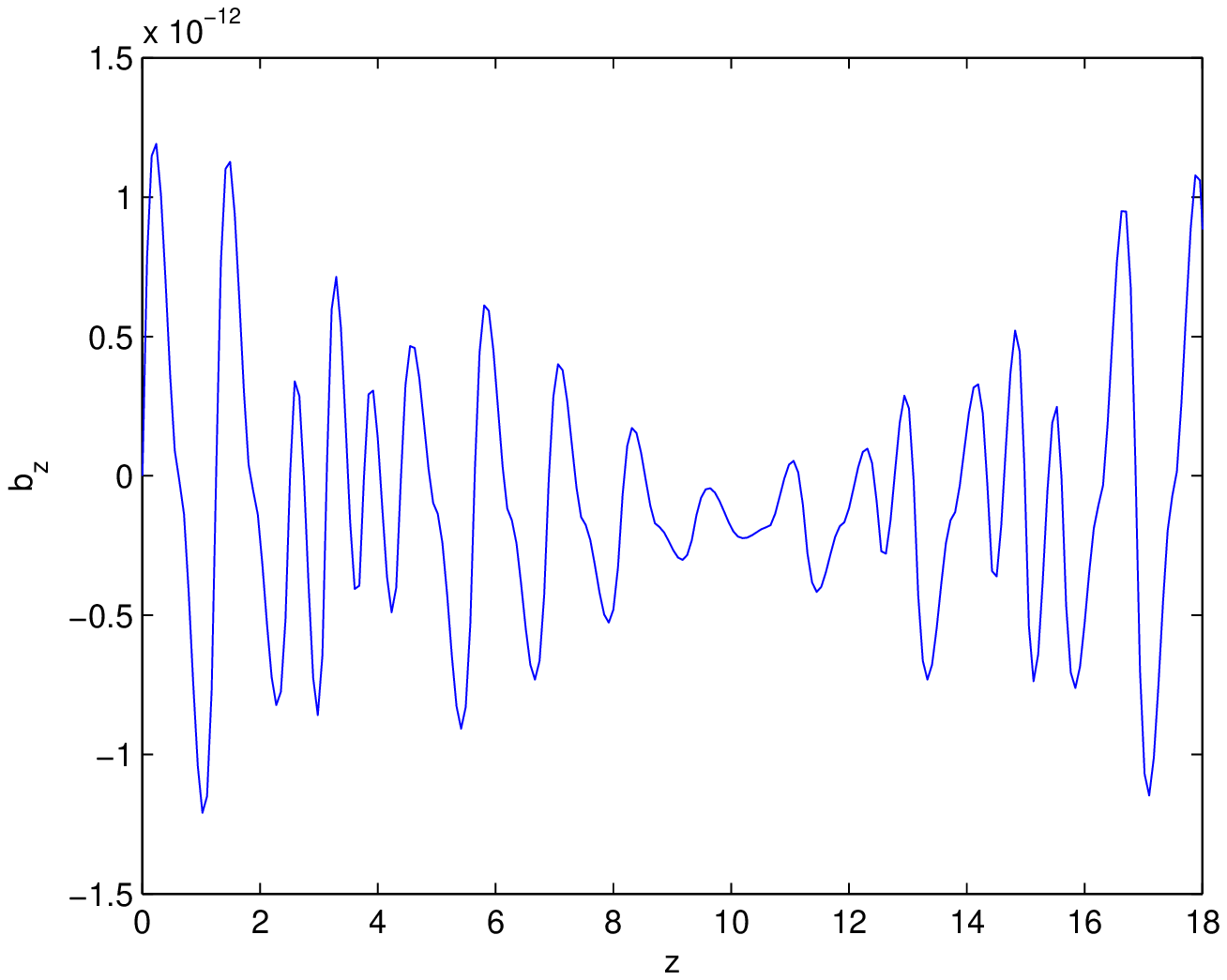}
\includegraphics[width=8cm]{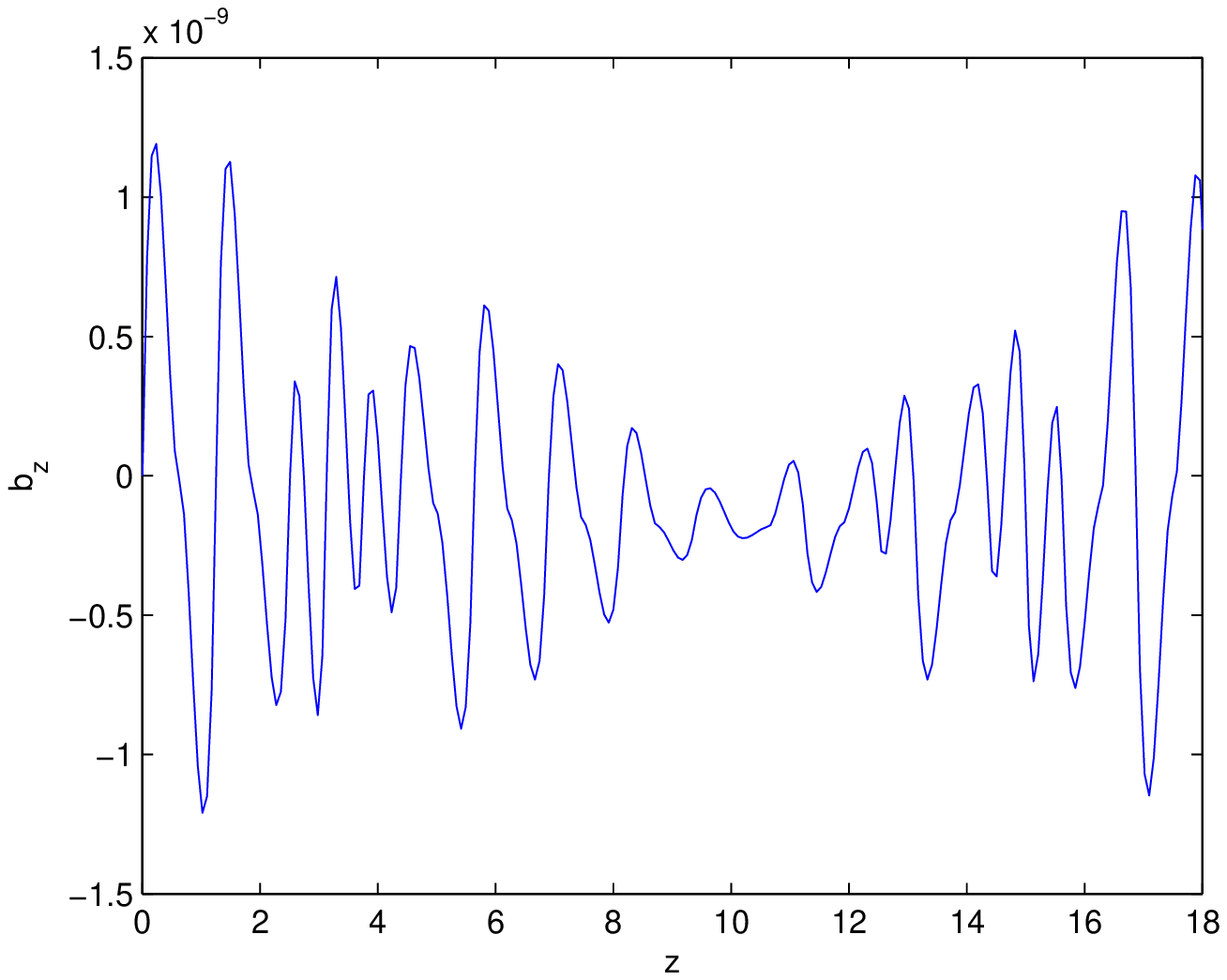}
\caption{The perturbed magnetic field ($b_{z}$) variations are showed with respect to $z$. Upper panel shows the kink mode along the
spicule axis. Lower panel shows the Alfv\'{e}n mode in the inhomogeneous layer. The perturbed magnetic
fields are normalized to $B_{0}$. \label{fig5}}
\end{figure}

\begin{figure}
\centering
\includegraphics[width=8cm]{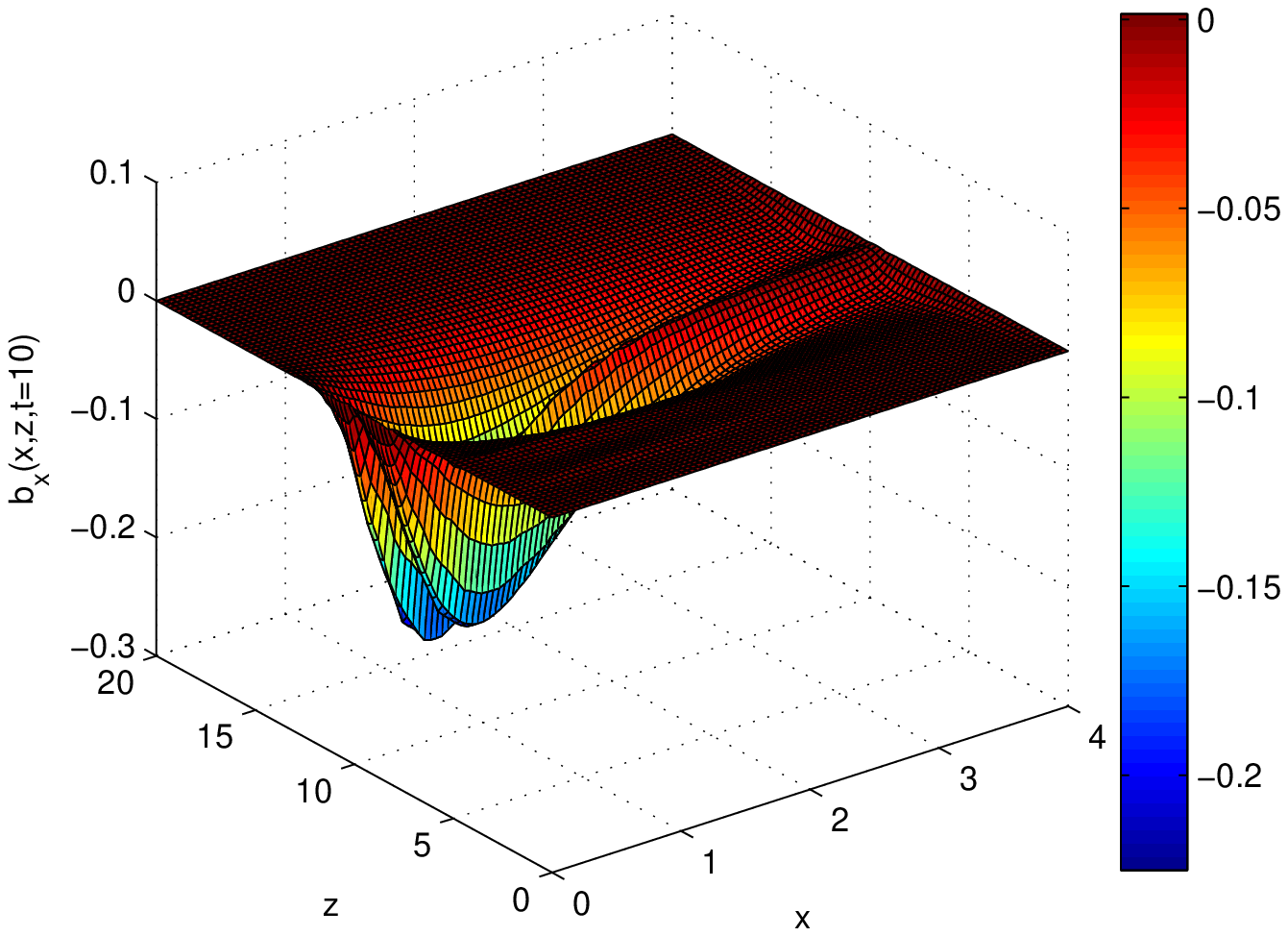}
\includegraphics[width=8cm]{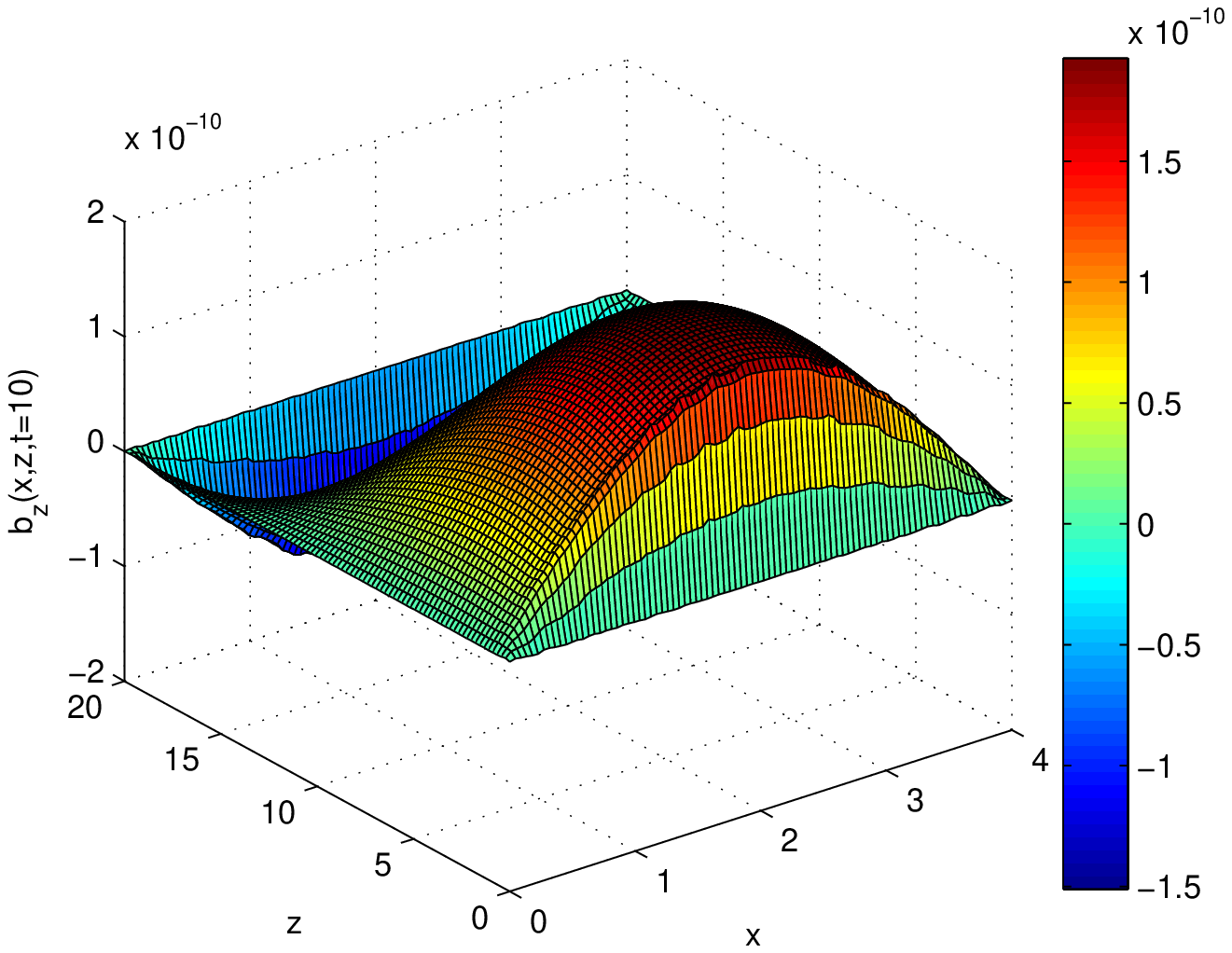}
\caption{The perturbed magnetic field is shown in $x-z$ space at $t=10 \tau$~s. \label{fig6}}
\end{figure}

\begin{figure}
\centering
\includegraphics[width=8cm]{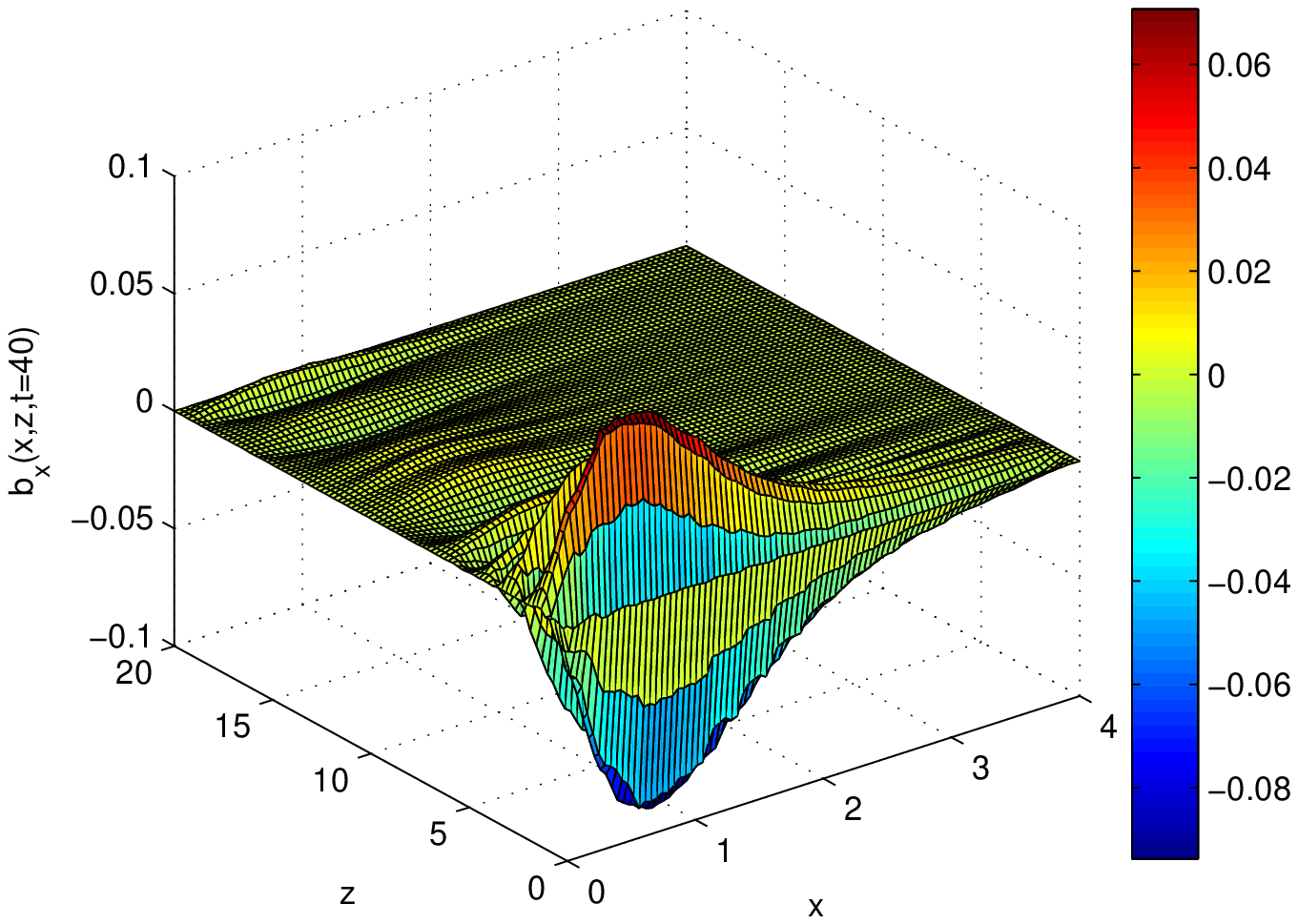}
\includegraphics[width=8cm]{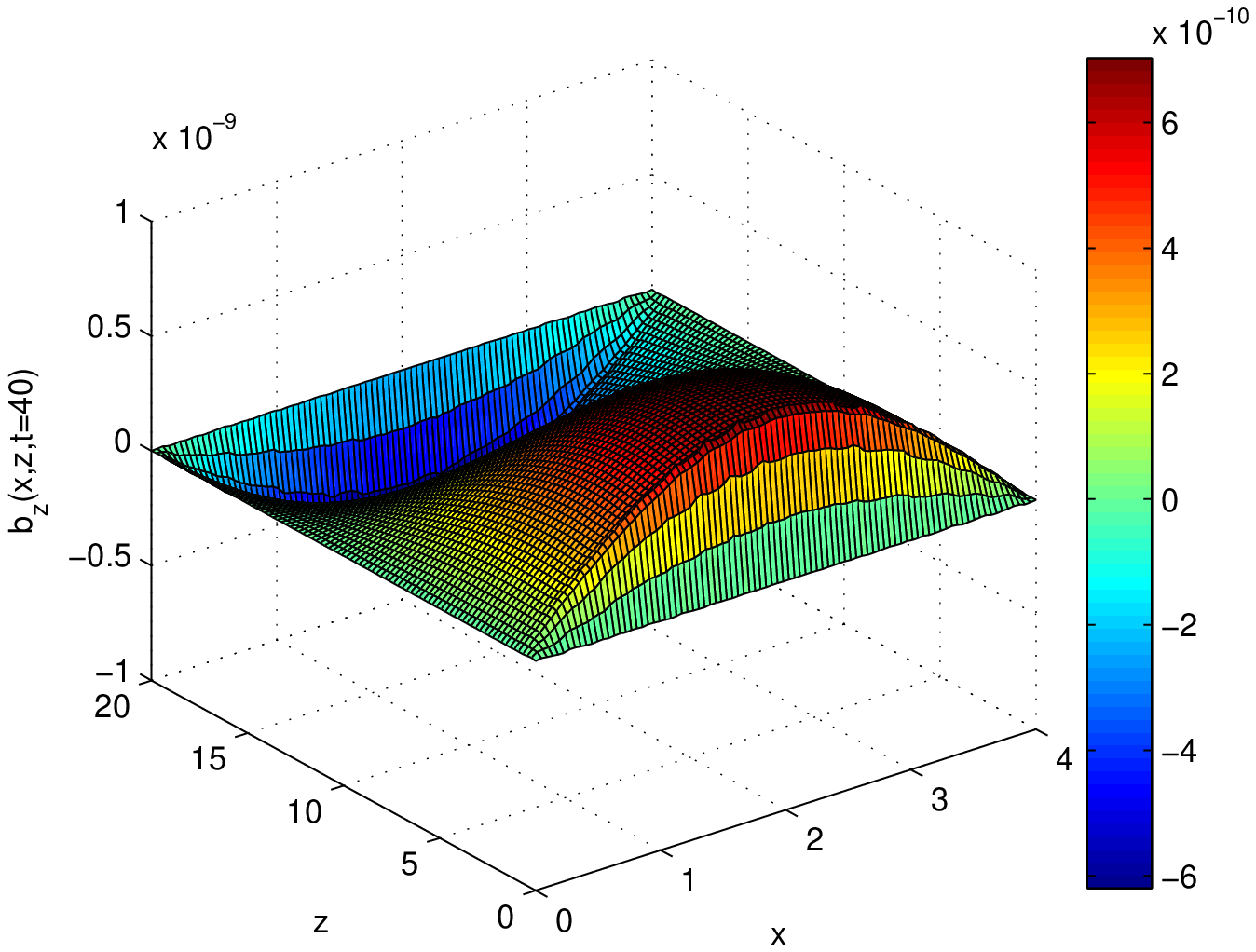}
\caption{The perturbed magnetic field is shown in $x-z$ space at $t=40 \tau$~s. \label{fig7}}
\end{figure}

\begin{figure}
\centering
\includegraphics[width=8cm]{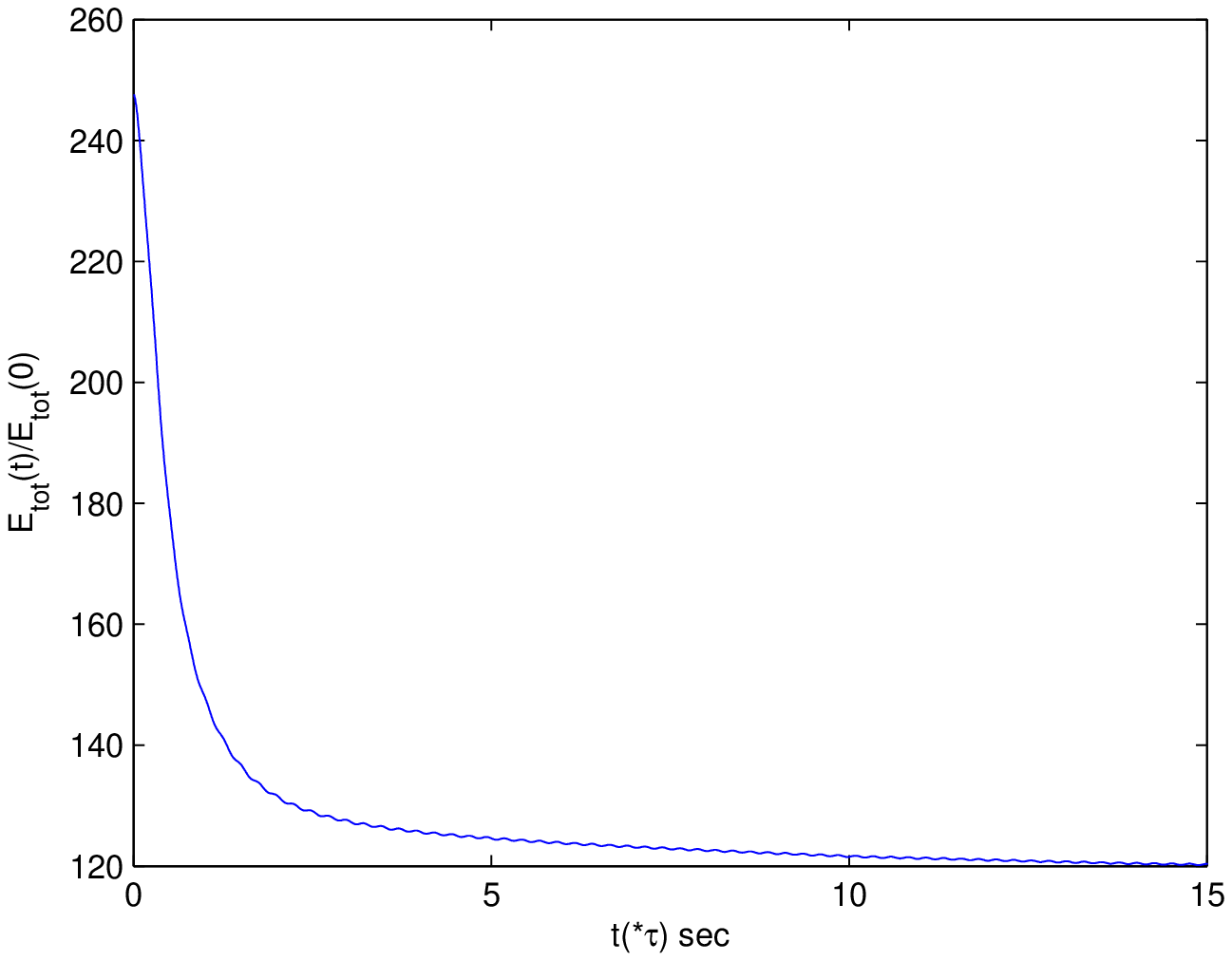}
\caption{Time variations of the total energy are presented. \label{fig8}}
\end{figure}

Figure~\ref{fig5} demonstrates variations of the perturbed magnetic field component, $b_{z}$, with respect to $z$ at the same locations as
in Figure~\ref{fig4}. The component $b_{z}$ has small fluctuations which indicates
an almost incompressible kink mode in the inhomogeneous layer (with an amplitude about $10^{-9}$).
\\The $3D$ plots of the perturbed magnetic field components with respect to $x$, $z$ are presented in Figures~\ref{fig6} and \ref{fig7}
at $t= 10 \tau$~s and $t= 40 \tau$~s, respectively. In these figures, also the spatial damping of the oscillations is seen along the tube axis.
In Figure~\ref{fig8}, the total energy normalized to the initial total energy is presented.
The wave energy is defined as $E_{tot}= 1/2(\rho(v_{x}^2+v_{y}^2+v_{z}^2) +1/\mu(b_{x}^2+b_{y}^2+b_{z}^2))$.
The coupling of kink mode to the local Alfv\'{e}n mode leads to a decrease in the kink wave energy in spicule which
shows the damping of the tube oscillations. The amplitude of total energy decreases exponentially with time.
The period of the kink waves ($P_{k}= 2L/C_{k}= 241.2$~s) are in good agreement with spicule lifetimes ($5-15$~min)
(Zaqarashvili \& Erd\'{e}lyi~\cite{Tem2009}).

\section{Conclusion}
\label{sect:discussion}
In this paper, it is attempted to investigate the behavior of magnetohydrodynamics waves observed in solar
spicules. The purpose is the study of their damping process by mode coupling mechanism. To do this, a typical
spicule is considered as a thin flux tube in the $x-z$ plane, and an initial perturbation is assumed as a
transverse kink wave in the lower z-boundary of the tube. The initial perturbation is applied on the velocity
and the magnetic field of the spicule, which it leads to transverse displacement of the spicule axis. These
transversal waves undergo a damping due to plasma inhomogeneity as height increases in which an energy
transferring takes place from the kink mode to the local Alfv\'{e}n mode. The amplitude of the Alfv\'{e}n
wave grows along the tube axis, and then experiences a spatial damping in the spicule. We observed that
total energy of the coupled kink to Alfv\'{e}n waves decreases with time exponentially.
\\In the approach of a thin tube with a thin inhomogeneous layer for a modeled spicule, the dissipation takes
place in this layer around $V_{A} = C_{k}$ where $C_{k}=\surd(\frac{1}{1+\rho_{e}/\rho_{0}})V_{A0}$
(Kukhianidze et al.~\cite{Kukh2006}). The exponentially damping of the spicule oscillations, gives
a damping time which it depends on the period of oscillation and the spicule parameters Pascoe et
al.~\cite{Pascoe2012}. This is obtained $\tau_{damp} = 8.5$~min in good agreement with the spicule
life time. This simulation is based on observational results reported by Ebadi et al.~\cite{Ebadi2012}.
They have studied these oscillations observationally and theoretically. They analyzed the time series
of Ca II H-line obtained from Hinode/SOT on the solar limb. The time distance analysis shows that the
axis of spicule undergos quasi-periodic transverse displacement at different heights from the photosphere.
The theoretical analysis also showed that the observed oscillations may correspond to the fundamental
harmonic of standing kink waves. In this case, our initial perturbation corresponds to some general photospheric
motion and our kink waves corresponds to the transverse oscillations in spicule axis observed from Hinode/SOT
by Tsuneta et al.~\cite{Tsuneta2008}.




\end{document}